\documentstyle[11pt,aaspp4]{article}

\begin{document}

\title{\baselineskip .5in \LARGE 
SCIENTIFIC REPORT ON LOW ENERGY ABSORPTION FEATURES DETECTED 
BY {\it Beppo}SAX LECS IN BRIGHT BLAZAR TARGETS }

\vskip 1.0in 

\LARGE{
\author{\huge Giovanni Fossati\footnote{fossati@sissa.it}}
\affil{International School for Advanced Studies (SISSA/ISAS), \\ 
via Beirut 4, 34014 Trieste, Italy}
\author{\huge Francesco Haardt\footnote{haardt@uni.mi.astro.it}}
\affil{Dipartimento di Fisica, Universit\`a degli Studi di Milano, \\ 
via Celoria 16, 20121 Milano, Italy}

\vskip 7 truecm
\centerline{\tt Ref. SISSA 146/97/A, August 1997} 

}

\vfill\eject
\normalsize{

\title{SCIENTIFIC REPORT ON LOW ENERGY ABSORPTION FEATURES DETECTED 
BY {\it Beppo}SAX LECS IN BRIGHT BLAZAR TARGETS}

\author{Giovanni Fossati}
\affil{International School for Advanced Studies (SISSA/ISAS), via
Beirut 4, 34014 Trieste, Italy}
\author{Francesco Haardt}
\affil{Dipartimento di Fisica, Universit\`a degli Studi di Milano, via
Celoria 16, 20121 Milano, Italy}

\section{Definition of the Problem}
\noindent
The apparency of low energy absorption features in Blazar LECS spectra 
has been reported during the early phase of the {\it Beppo}SAX mission
by Grandi et al. (1997) for 3C~273 Science Verification Phase (SVP) and 
by Giommi et al. (1997) for PKS 2155--304 SVP, with data reduction
and analysis tools released prior to July 1997.
Grandi et al. reported a broad absorption feature
centered at $0.53\pm 0.07$ keV (hereinafter energies are measured in 
the observer frame), while the LECS data of PKS 2155--304 show a 
complex pattern of absorption edges. 
A fit with up to three edges (at energies 0.2, 0.5, 0.7 keV) seems to be
consistent with the data. 
In LECS Core Program (CP) data of 3C~273 we also detected an absorption
edge at $0.7\pm 0.1$ keV with $\tau=0.33\pm 0.14$ (Haardt et al. in prep), 
not consistent with that observed in SVP by Grandi and co--workers.

\section{Test Procedure}
\noindent
In order to investigate the reliability of our edge detection, we 
decided to perform a test aimed at the discovery of possible systematic 
errors in the LECS response matrix.
In the analysis presented in this report we refer to LECS/MECS calibration
files (effective area, response matrix) and analysis software ({\tt SAXDAS}) 
released by {\it Beppo}SAX Science Data Center (SDC) {\bf prior to July 1997}.  
The test was performed as follows: 

\begin{enumerate}
\item{}{we defined for seven LECS datasets of bright blazars 
(Mkn 421 May 2$^{\rm nd}$ and 4$^{\rm th}$ 1997 pointings, PKS 2155--304 SVP, 
Mkn 501 April 7$^{\rm th}$ and 16$^{\rm th}$ 1997 pointings, 3C~273 summed
over CP observations, 3C~273 SVP) the "best continuum", i.e. LECS data were
fitted by a single or broken power law absorbed by Galactic column density
({\tt wabs bknpower} model in {\tt XSPEC 9.0}); 
prior to fitting, LECS data were rebinned according to {\tt grouping2} 
template as provided by {\it Beppo}SAX SDC (see at {\tt 
ftp://www.sdc.asi.it/pub/sax/cal/responses/grouping/}). 
The high energy branch of {\tt bknpower} was fixed at the value obtained 
fitting the MECS data;} 
\item{}{each data/model ratio obtained with the so defined ``best
continuum" was then divided by a ratio averaged over 
{\it the remaining 6 datasets}, i.e. excluding from the averaged ratio the 
ratio corresponding to the dataset under study. 
In this way, should a particular feature be present only in 
the dataset under study, it would be conserved going through the process 
of "average over the rest". 
On the other hand, features common to all the datasets would be smeared out. 
Note that the averaged ratios were obtained performing an arithmetic mean 
of the individual ratios, with quadratic propagation of errors. 
No weighted mean was used whatsoever. 
This is because we wanted to avoid the averaged ratio to be led by the best 
S/N sources, such as PKS 2155--304}.
\end{enumerate}

\section{Results}
In Figure 1, left column, 
the data/model ratios for the 7 datasets are presented. 
The upper left panel is the ratio averaged over all the 7 datasets 
(note that this average ratio is shown for mere
clarity purposes, but it {\it was not} actually used in the process, as it
contains information from {\it all} the 7 datasets). 
In the right column the same ratios are divided by the relevant 
{\it ``average over the rest"} ratios, as discussed above. 
It can be easily noticed that all the "corrected" ratios but one are 
consistent with a constant=1, without any significant deviation. 
The only case which seems to be not affected by the average procedure is 
3C~273 SVP, which shows the absorption feature as observed 
in the original data. 

To be more quantitative we used the {\it "average of the rest"} ratios to
modify the PHA file of each dataset, using the {\tt fcreate} tool in {\tt
FTOOLS}. 
In this way we were able to obtain seven "corrected" PHA files ready to be 
used by {\tt XSPEC}. 
We confirm the results of the visual inspection of the ratios discussed above: 
all the datasets (but 3C~273 SVP) can be fitted by (pure continuum) model 
{\tt wabs bknpower} (i.e. absorbed broken power law) without need of extra
features. 
The case of PKS 2155--304 is shown in Fig.~2. The upper panel shows the ratio 
of the {\tt wabs bknpower} model to the data prior correction, the middle panel 
the ratio of {\tt wabs bknpower edge} to the same original data, and the
lower panel the ratio of a simple {\tt wabs bknpower} model to the
corrected data. 

As said above, only 3C~273 SVP retains an evident absorption feature, 
essentially at the same energy and of the same extent of what was observed 
in the original uncorrected data (see Figure~3).

We finally checked that results are indeed identical if we propagate in
the fitted data the errors of the averaged ratios, rather 
than simply sticking with the Poissonian errors of the original data.

\section{Discussion}
\noindent 
We set up a robust method to investigate the reliability of absorption 
features detections in LECS data of bright blazars. 
The method has been applied to 7 datasets. 
The analysis shows that similar features are common to all datasets 
(see next for 3C~273 SVP).
This suggests that most of the features observed from 0.2 to 1 keV are possibly
due to miscalibration of the instrument response, as of present date. 
The large deviations below 0.5 are probably due to the Carbon edge (at
0.29 keV), which can affect the residuals up to 1 keV. 
Of course a different interpretation of our results is still possible, 
namely that the common feature is indeed real. 
Should be that, interesting problems in the physical interpretation of
blazar sources would be raised. 

As discussed above, only 3C~273 SVP still shows a clear absorption feature 
after correction. The fact that the correction does not improve the 
$\chi^2$ when fits are performed to the same
model means that the feature detected in this dataset is "unique", 
in the sense that is not shared by any of the other datasets considered. 
One may argue that, if the "common" feature is indeed a miscalibration 
product, it should be apparent also in 3C~273 SVP. We note however  that 
the "unique" absorption feature is probably deep enough to overwhelm the 
"common" feature.   

\acknowledgements
We thank the SAX--Team for providing the data of PKS 2155--304 SVP, 
for private use and testing purposes, and P. Grandi for useful and 
fruitful discussions.

\vskip 1 truecm
\noindent
\fbox{\parbox{\textwidth}{{\bf Note added:}
the following release of LECS matrix and reduction pipeline (September
1997) has solved the problem, confirming our conclusion that the features
detected below 0.5 keV using the first software release were indeed of
instrumental origin (Guainazzi M., Grandi P., {\it Beppo}SAX SDC 
technical report -- TR-014, August 1997; A.~Orr. et al., SDC technical
report -- TR-015 August 1997).}}

\newpage

\begin{figure}
\plotone{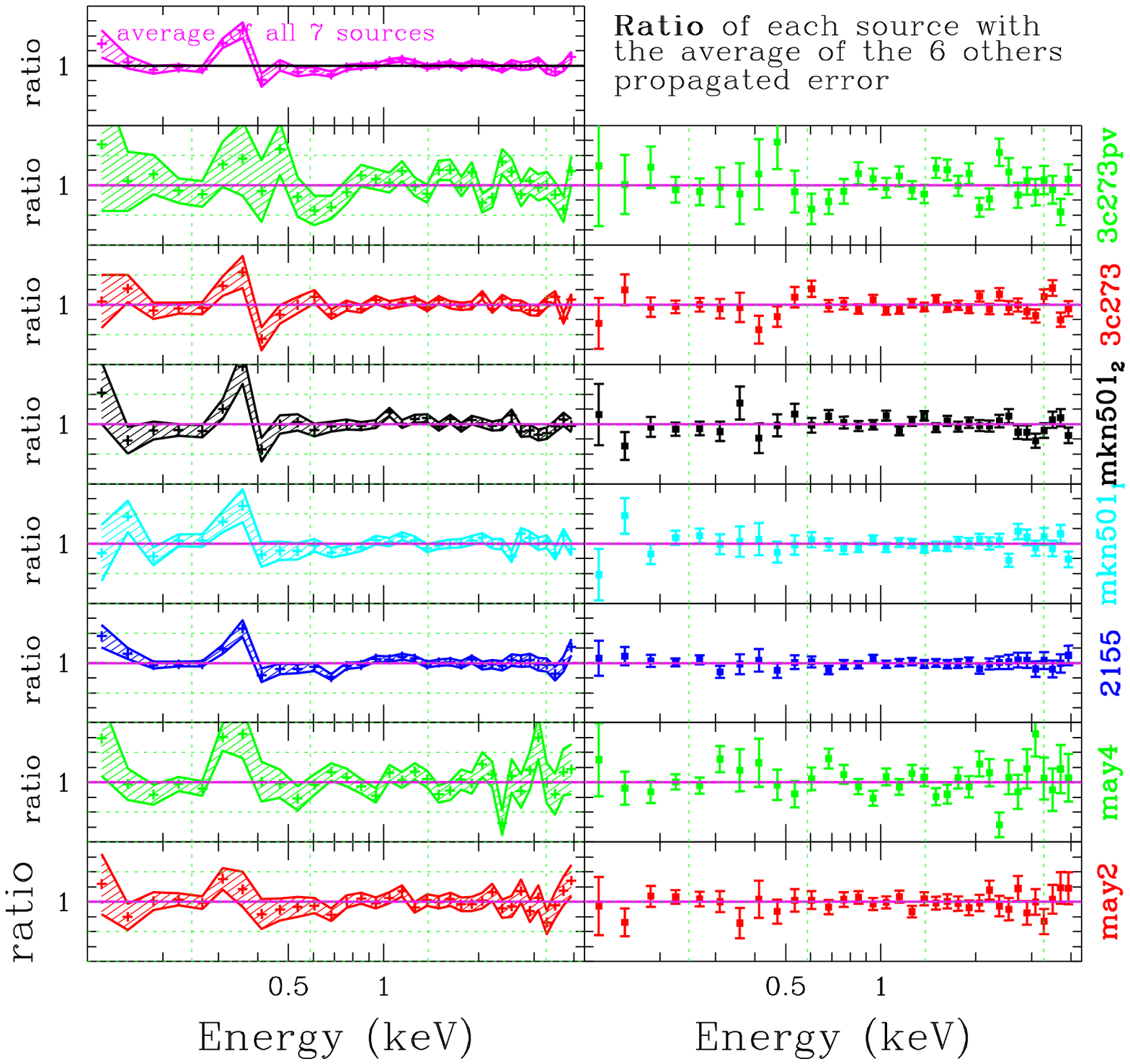}
\caption{Left column: data/model ratios for the seven datasets. Upper left 
panel is the average of the seven. Right column: data/model ratios divided 
by the "average of the rest" ratio. See text for details.}
\end{figure}

\begin{figure}
\plotone{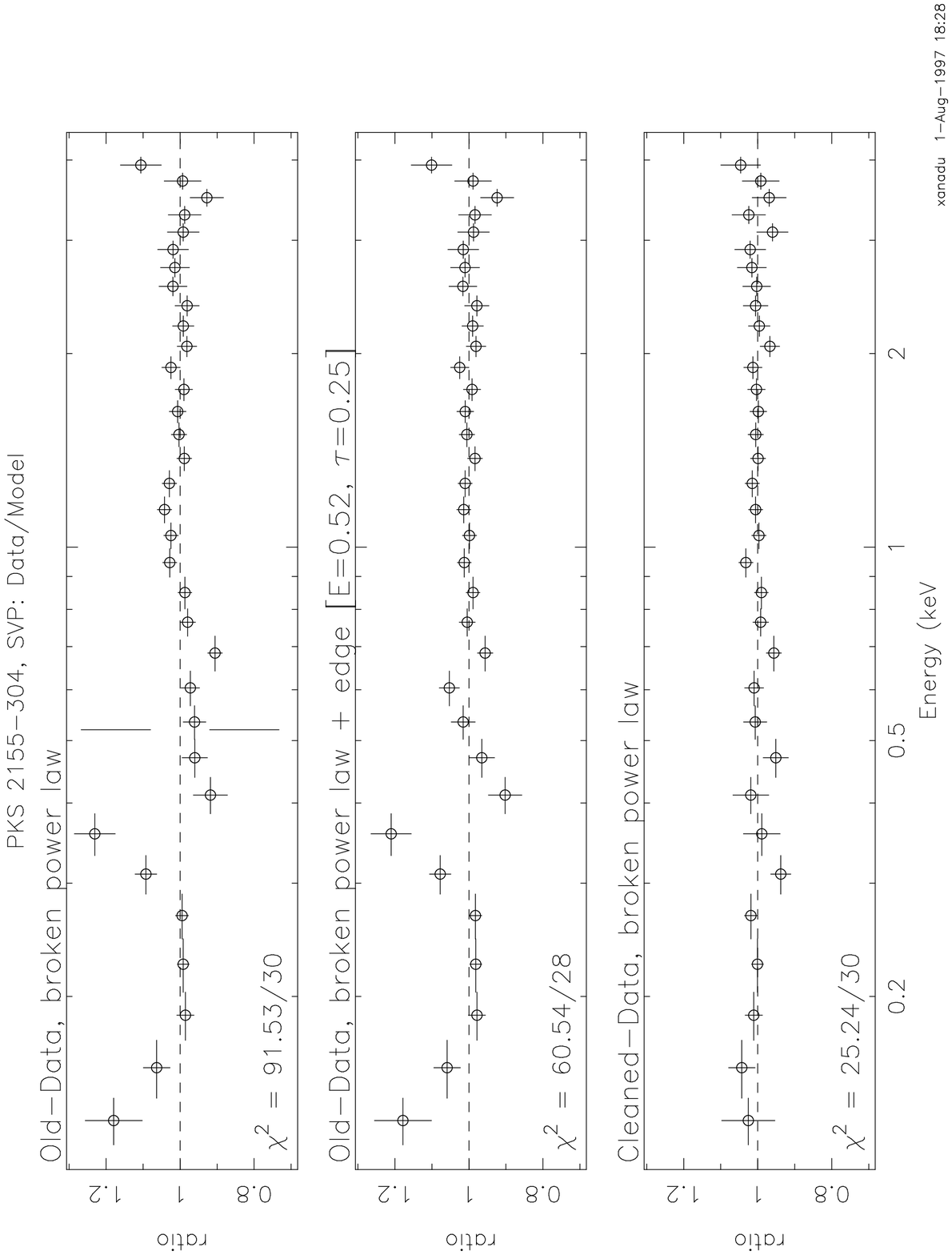}          
\caption{Residuals of different models to old and cleaned PKS 2155 LECS 
data. See figure labels and text for details.}
\end{figure}

\begin{figure}
\plotone{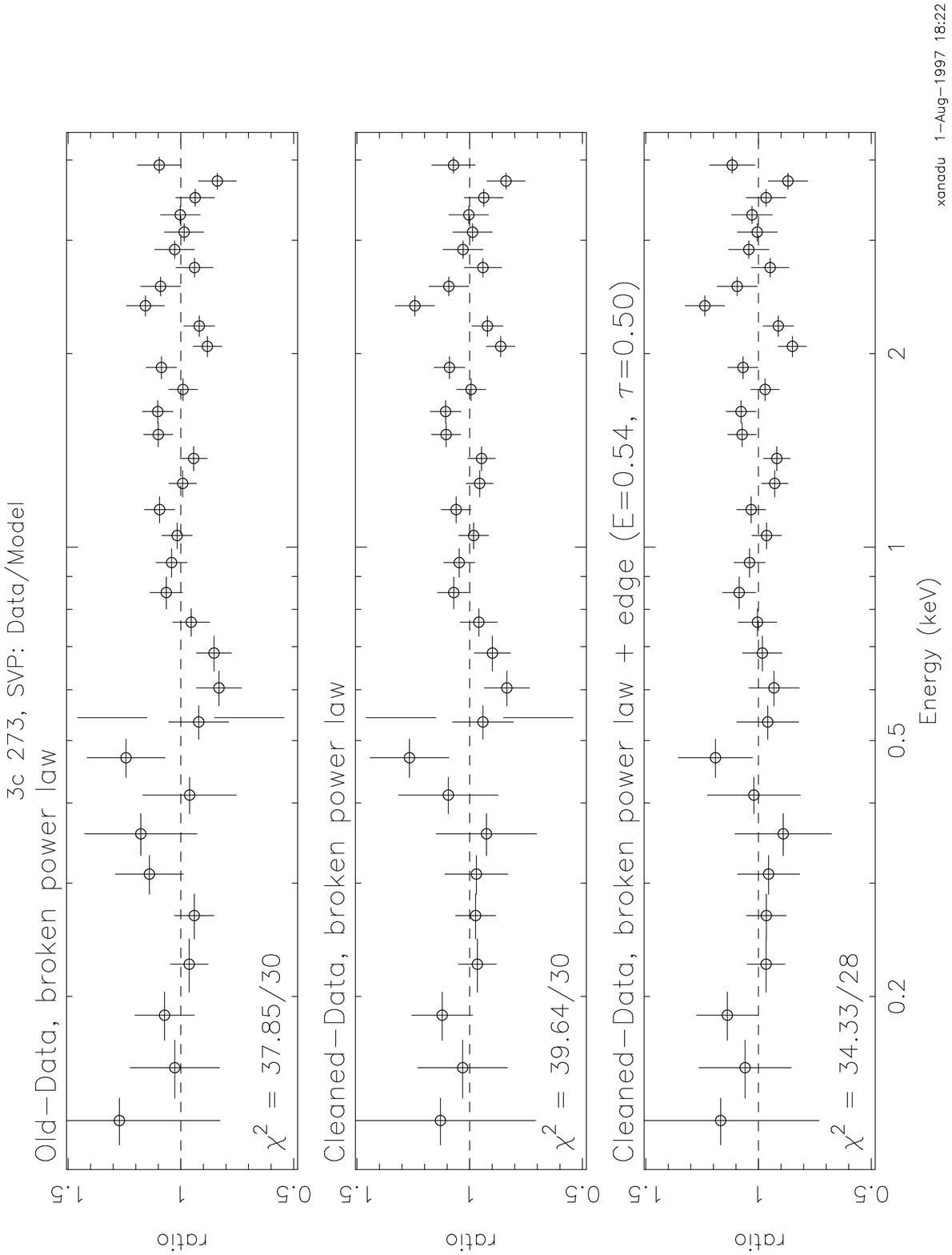}          
\caption{Residuals of different models to old and cleaned 3C~273 SVP LECS 
data. See figure labels and text for details.}
\end{figure}
}

\end{document}